\documentclass[%
reprint,
amsmath,amssymb,
aip,
floatfix,
]{revtex4-1}
\usepackage{amsmath}
\usepackage{graphicx}% Include figure files
\usepackage{dcolumn}% Align table columns on decimal point
\usepackage{csquotes}
\usepackage{bm}% bold math
\usepackage{float}
\usepackage{color}
\usepackage[colorlinks=true, allcolors=blue]{hyperref}
\usepackage[version=4]{mhchem}

\usepackage{subcaption}
\graphicspath{ {./figures/} }

\begin{document}

%\preprint{APS/123-QED}
\title{Excess Density as a Descriptor for Electrolyte Solvent Design}

%%%%%%%%%%%%%%%%%%%%%%%%%AUTHORS%%%%%%%%%%%%%%%%%%%%%%%%%%%%%%%%%%%
\author{Celia Kelly}
\affiliation{%
Department of Mechanical Engineering, Carnegie Mellon University, Pittsburgh, Pennsylvania 15213
}

\author{Emil Annevelink}
\affiliation{%
Department of Mechanical Engineering, Carnegie Mellon University, Pittsburgh, Pennsylvania 15213
}

\author{Adarsh Dave}
\affiliation{%
Department of Mechanical Engineering, Carnegie Mellon University, Pittsburgh, Pennsylvania 15213
}

\author{Venkatasubramanian Viswanathan}%
\email{venkvis@umich.edu}
\affiliation{%
Department of Mechanical Engineering, Carnegie Mellon University, Pittsburgh, Pennsylvania 15213
}
\affiliation{Department of Mechanical Engineering, University of Michigan, Ann Arbor, MI 48109}
\affiliation{Department of Aerospace Engineering, University of Michigan, Ann Arbor, MI 48109}

\date{\today}

\begin{abstract}

Electrolytes mediate interactions between the cathode and anode and determine performance characteristics of batteries. Mixtures of multiple solvents are often used in electrolytes to achieve desired properties, such as viscosity, dielectric constant, boiling point, and melting point. Conventionally, multi-component electrolyte properties are approximated with linear mixing, but in practice, significant deviations are observed. Excess quantities can provide insights into the molecular behavior of the mixture and could form the basis for designing high-performance electrolytes. Here we investigate the excess density of commonly used Li-ion battery solvents such as cyclic carbonates, linear carbonates, ethers, and nitriles with molecular dynamics simulations. We additionally investigate electrolytes consisting of these solvents and a salt. The results smoothly vary with mole percent and are fit to permutation-invariant Redlich-Kister polynomials. Mixtures of similar solvents, such as cyclic-cyclic carbonate mixtures, tend to have excess properties that are lower in magnitude compared to mixtures of dissimilar substances, such as carbonate-nitrile mixtures. We perform experimental testing using our robotic test stand, Clio, to provide validation to the observed simulation trends. We quantify the structure similarity using SOAP fingerprints to create a descriptor for excess density, enabling the design of electrolyte properties. To a first approximation, this will allow us to estimate the deviation of a mixture from ideal behavior based solely upon the structural dissimilarity of the components. 

\end{abstract}

\maketitle

\section{Introduction}
Electrolytes are a key component of lithium-ion batteries and contribute to several aspects of battery performance, such as safety, charge rate, power output, and cycle life.\cite{KangElectrolyteReview,KangReview} As such, there are many properties that relate to electrolyte performance, such as ionic conductivity, thermal stability, and electrochemical stability.\cite{KangElectrolyteReview, KangReview,wang2015development} The electrolyte should satisfy many requirements simultaneously, such as having high Coulombic efficiency, maximizing specific energy, having good oxidative stability, increasing cost-effectiveness, and meeting safety and processability requirements.\cite{yu2020molecular} The solvent in particular should dissolve salts, enable ion transport, be inert to cell components such as the anode and cathode, remain liquid in a large temperature range, and satisfy safety requirements.\cite{KangElectrolyteReview} A mixture of different components is often used for the liquid electrolyte in order to take advantage of the desirable properties of each component, such as the high dielectric permittivity of ethylene carbonate and the low viscosity of linear carbonates such as diethyl carbonate or dimethyl carbonate. \cite{KangReview,xu2021li} Understanding the properties of such mixtures is important for the development of improved liquid electrolytes.\cite{oldiges2018understanding} Molecular interactions can provide insight into the performance of an mixture, thereby enabling the design of improved battery electrolytes.\cite{yu2020molecular}

Excess properties (e.g. excess density, excess molar volume) have been used as a method of understanding the behavior of mixtures \cite{renon1968local}. Excess density can be calculated as the deviation of a mixture from ideal mixing behavior.\cite{shukla1988thermodynamic,kamble2012study,navarkhele2022dielectric} Excess density has been used as an index for repulsion and attraction in mixed liquid electrolytes consisting of solvate ionic liquids and 1,1,2,2-tetrafluoroethyl 2,2,3,3-tetrafluoropropyl ether (HFE).\cite{takahashi2019physicochemical} Excess density of crude oil and brine mixtures were suggested to be caused by a phase separation and were further correlated to the kinematic viscosity of oil.\cite{density_wateroil} Another work examined excess molar volume, excess polarity, and excess refractive index of binary mixtures of room temperature ionic liquids and stated that the small excess values suggested that the properties of the mixture could be easily tuned by mixing components. \cite{seki2019densities} The behavior of excess molar volume has been associated with both geometric factors and intermolecular forces,\cite{rafiee2016study} and has been stated to rely on the differing strengths of the contractive and expansive factors.\cite{saleh2005excess} Negative contributions to excess molar volume include intermolecular interactions such as H-bonding and interstitial accommodation of two different molecules.\cite{moosavi2017densities} Positive contributions to excess molar volume include dissociation of similar molecules on mixing.\cite{moosavi2017densities} Additionally, the Redlich-Kister coefficients for excess molar volume can be used to calculate partial molar volume at infinite dilution.\cite{hartono2009density} 

Excess properties are frequently fit with Redlich-Kister polynomials.\cite{tomiska1984mathematical,kamble2012study,moosavi2017densities} Redlich-Kister polynomials were originally developed for the representation of the thermodynamic properties of binary solutions.\cite{redlich1948algebraic} They are commonly used to correlate excess properties as functions of mole fraction.\cite{hartono2009density} The Redlich-Kister polynomial can show the sign and magnitude of non-ideality; the magnitude of the first coefficient will reflect the amount of deviation from linear mixing, while the sign of the first coefficient will the reflect the direction of the deviation from linear mixing. Thus, the Redlich-Kister polynomial, fit to excess properties, can show non-ideality in the mixture. 

Both experiments and simulations can be used to calculate excess densities of a mixture of two substances, by either measuring or calculating the density of the pure substances and of the mixture of the substances.\cite{wagner2004densimeters,wang2011application} Experimentally, the density of a substance can be found by measuring the mass of a given volume of a substance with a mass balance, and then dividing the mass by the volume.\cite{dave2022autonomous} Alternatively, density could be measured directly.\cite{wagner2004densimeters,takahashi2019physicochemical} The excess density can then be calculated from the density as described in the Methodology section. However, this requires the time, facilities, equipment, and substances to run these experiments. Although they require specific parameters to approximate the forces acting on the molecules and require computational resources, molecular dynamics simulations can be used to investigate properties which are difficult to research experimentally and can be used to guide experimental work.\cite{hollingsworth2018molecular,wang2011application2,liu2018molecular} By modeling liquid electrolytes, we can use molecular dynamics simulations to reveal the molecular origins behind the behavior of an electrolyte.\cite{fang2023elucidating} 

In this work, we will discuss the results of molecular dynamics simulations of several solvents commonly used in liquid electrolytes for Li-ion batteries and simulations of these solvents with a salt. The excess density was calculated from the density of the simulated mixtures. To capture trends in the excess density, the similarity of the molecules was  quantified. It was observed that the magnitude of the maximum excess density of the mixture tended to increase as the similarity of the molecules decreased. As the magnitude of excess density is linked to intermolecular interactions, this indicates strong interactions between molecules that are dissimilar and provides a quantitative basis for electrolyte mixture design.

\section{Methodology}

\begin{figure}[htb!]
\centering
\includegraphics[width=\columnwidth]{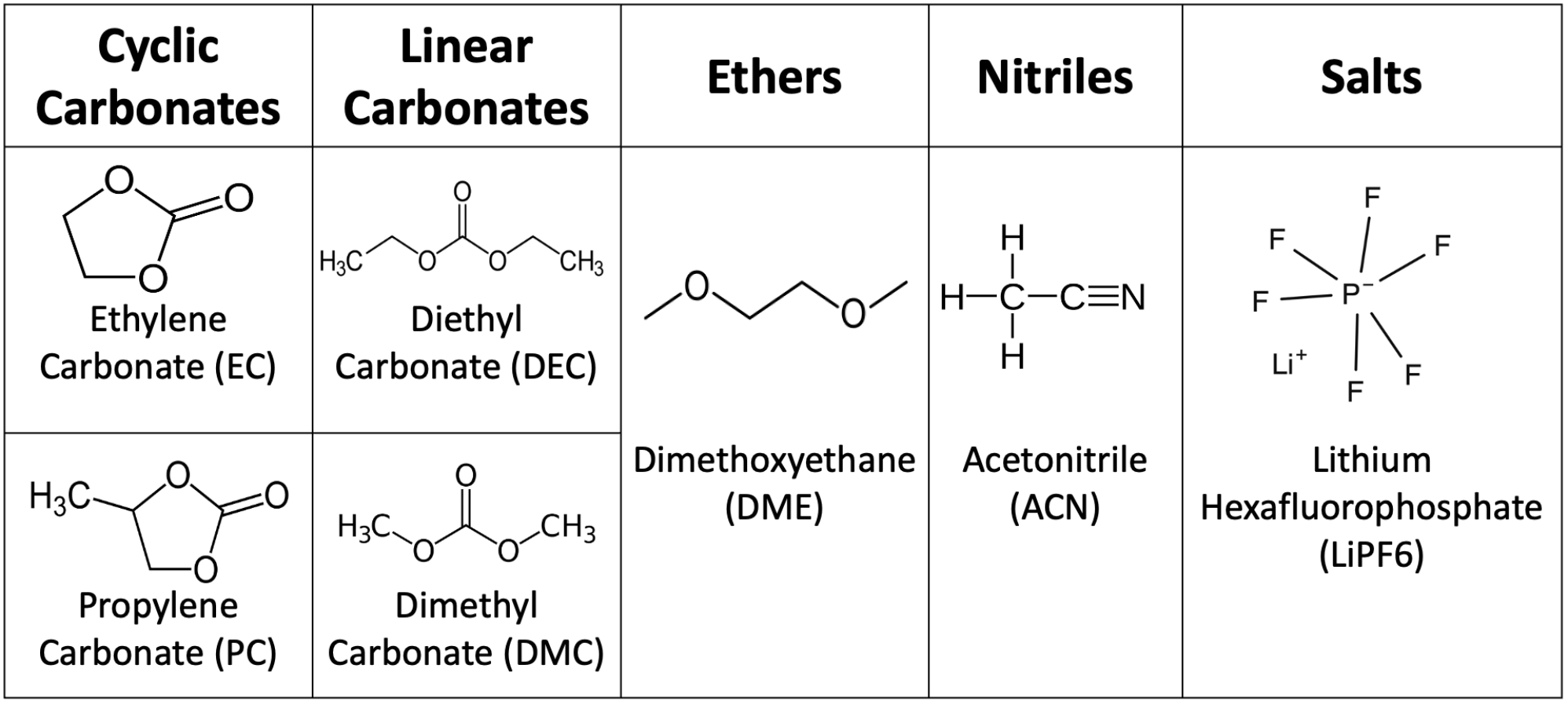}
\caption{Solvents used in molecular dynamics simulations.}
\label{fig:solvents}
\end{figure}

The density of common solvents is determined by performing molecular dynamics simulations of solvents and binary mixtures of solvents. The solvents, shown in Figure \ref{fig:solvents}, include cyclic carbonates, linear carbonates, ethers, and nitriles. The mixtures used include cyclic-cyclic carbonate mixtures, cyclic-linear carbonate mixtures, carbonate-ether mixtures, carbonate-nitrile mixtures, and ether-nitrile mixtures. 

GROMACS molecular dynamics simulation software was used for all simulations.\cite{berendsen1995gromacs,abraham2015gromacs} When setting up the molecular dynamics simulations, 1000 molecules were randomly inserted into a simulation box using the gmx insert-molecule function available in GROMACS.\cite{gromacsmanual} For mixtures, molecules were inserted in three ratios: 1:3, 1:1, and 3:1. In the case of the solvent and salt mixtures, 10 lithium ions and 10 hexafluorophosphate ions were added to the simulation box, for a ratio of 1:100 salt molecules to solvent molecules. 

\begin{figure*}[htb!]
\centering
    \includegraphics[width=\textwidth]{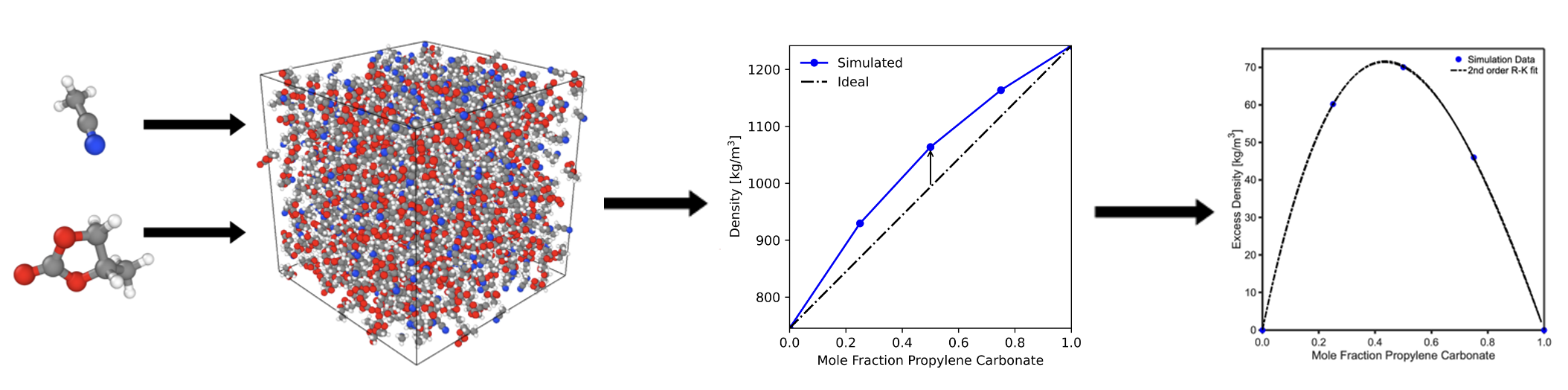}
\caption{Two molecules, shown here as acetonitrile and propylene carbonate, are inserted into the simulation box to create a mixture. From the resulting simulation box, the density of the mixture is found. The difference between the ideal density according to linear mixing and the calculated density is the excess density. The excess density is then fit to a Redlich-Kister polynomial.}
\label{fig:excess_linear}
\end{figure*}

Once the simulation was set up, energy minimization was performed using steepest descent minimization for a maximum of 50,000 steps. Following energy minimization, the system was equilibrated using the NVT ensemble, with a Nose-Hoover thermostat set to 313 K for mixtures including ethylene carbonate and 300 K for all other mixtures. A higher temperature was chosen for ethylene carbonate mixtures because EC has a melting point of 36 $^{\circ}$C or approximately 309 K and as such is solid at 300 K, whereas the other solvents have melting points of -4.9 $^{\circ}$C (PC), -68 $^{\circ}$C (DME), -44 $^{\circ}$C (ACN), -43 $^{\circ}$C (DEC), and 4 $^{\circ}$C (DMC) and are liquid at 300 K.\cite{stenutzn} 40 $^{\circ}$C or 313 K is sometimes used in literature for simulations with EC.\cite{borodin2006litfsi}

NVT equilibration was performed for 500 ps before moving on to NPT equilibration. NPT equilibration was performed for 1 ns, keeping thermostat settings the same and using a Parinello-Rahman barostat set to 1 bar. Following the equilibration stages, a production run was performed in the same NPT ensemble for 10 ns. The average density of the simulation was extracted from the production run using the gmx energy function available in GROMACS.\cite{gromacsmanual}

The OPLS-AA force field was used for all simulations. The functional form of the force field expresses the potential energy as a combination of the energy from bond stretching, angle bending, torsion from dihedrals, and non-bonded terms:\cite{oplsaa} 
\begin{equation}
\begin{split}
    E(\phi)=E_{bond}(\phi) + E_{angle}(\phi) \\ + E_{torsion}(\phi)  +E_{nonbonded}(\phi)
\end{split}
\end{equation}

The force field parameters for EC and DEC were obtained from the Virtual Chemistry database \cite{virtualchem_opls,virtualchem_opls2}. Parameters for ACN, DME, and DMC were obtained from our previous work.\cite{khetan2018understanding, zhang2022predicting} Parameters for PC were taken from You et al.\cite{you2016dielectric} Parameters for PF6 were taken from the work of Doherty et al.\cite{doherty2017revisiting} Benchmarks were performed by comparing the density calculated from the simulation to reported densities for the pure solvents in order to ensure that the parameters found achieved decent agreement with the reported densities. 

After extracting the density of each mixture from the simulation, the excess density was calculated as follows: 
\begin{equation} \label{eq:excess_eq}
\rho_E=\rho_{mixture}-(\rho_1x_1+\rho_2x_2)
\end{equation}
where $\rho_{mixture}$ is the density of the mixture as measured from the simulation, $\rho_1$ is the density of the first solvent measured from the simulation,  $\rho_2$ is the density of the second solvent, $x_1$ is the mole fraction of the first solvent, and $x_2$ is the mole fraction of the second solvent. This is illustrated in Figure \ref{fig:excess_linear}. 

The resulting excess density values were then fit to a Redlich-Kister polynomial according to the equation,\cite{redlich1948algebraic} 
\begin{equation} \label{eq:RK_eq}
\rho_E = x(1 - x)[a_1 + a_2(2x - 1) + a_3(2x - 1)^2]
\end{equation}
in which $x$ is the mole fraction of one of the solvent components, and $a_1$, $a_2$, and $a_3$ are the resultant coefficients from fitting the results to the polynomial.

Radial distribution functions were additionally calculated for certain mixtures using the gmx rdf function available in GROMACS, referencing the center of mass of each molecule.\cite{gromacsmanual} The radial distribution function describes the distribution of particles around a reference particle. At small distances, the radial distribution function tends to 0 due to the strong repulsive forces at close range; at long distances, it tends to unity, which is the radial distribution function for an ideal gas. As such, any deviation from unity describes intermolecular interactions.\cite{FRENKEL200263,KirkwoodBoggs} In GROMACS, the radial distribution function of  particle B around reference particle A is defined according to the equation,\cite{gromacsmanual}
\begin{equation}
g_{AB} = \frac{\langle\rho_{B}(r)\rangle}{\langle\rho_{B}\rangle_{local}}
\end{equation}
where $\langle\rho_{B}(r)\rangle$ is the particle density of B around reference particle A at distance r, and $\langle\rho_{B}\rangle_{local}$ is the particle density of B averaged over all spheres around particles A with radius $r_{max}$.\cite{gromacsmanual}

In order to determine whether there are any observable trends in the excess densities depending on the similarity of the molecules, it is necessary to quantify the similarities. This was done using smooth overlap of atomic position (SOAP)  fingerprints and the regular entropy match (REMatch) kernel method.\cite{de2016comparing} SOAP fingerprints have been used to describe coordination environments in a way that is invariant to to translations, rotations and permutations of atoms.\cite{de2016comparing} Given two structures, an environment covariant matrix containing all pairings of environments can be computed. This matrix, which contains all information on pair-wise similarity, can then be used to create a global kernel to compare two structures. The REMatch method is one method of doing so.\cite{de2016comparing} The DScribe implementation of the SOAP fingerprint and REMatch kernel method was used for this paper.\cite{dscribe, dscribe2}

\section{Results}

\subsection{Benchmarks}

\begin{table}[htb!]
\centering
\resizebox{\columnwidth}{!}{\begin{tabular}{ |c|c|c|c| } 
 \hline
 Solvent & Simulation [$kg/m^3$] & Experiment [$kg/m^3$] & Percent Error\\
 \hline
 Ethylene Carbonate, 300K (EC) & 1322 &	1338$^a$  &	-1.20   \\ 
 \hline
 Ethylene Carbonate, 313K (EC) & 1308 &	1321$^b$  &	-0.98   \\ 
 \hline
 Propylene Carbonate (PC) & 1241  & 1201$^a$, 1200$^b$   & 3.33, 3.42  \\ 
 \hline
 Dimethyl Carbonate (DMC) & 1075  & 1073$^a$ 1063$^b$ & 0.19, 1.13   \\ 
 \hline
 Diethyl Carbonate (DEC) & 994  &	976$^a$, 969$^b$  &	1.84, 2.58   \\ 
 \hline
 Dimethoxyethane (DME) & 865  & 869$^a$  & -0.46  \\ 
 \hline
 Acetonitrile (ACN) & 745  & 782$^a$  &	-4.73   \\ 
 \hline
\end{tabular}}
\caption{Density values calculated from the simulation and experimental density values as reported in literature are shown with the percent error of density. $^a$Values taken from the homepage of R. Stenutz.\cite{stenutzn} $^b$Values taken from Seo et al.\cite{seo2015role} }
\label{table:benchmark}
\end{table}

To benchmark the simulations, first simulations of the pure substances were performed to compare the calculated density to reported experimental values. All simulations were performed at 300K and 1 bar. An additional benchmark was performed for ethylene carbonate at 313K. The results are shown in Table \ref{table:benchmark}. The largest difference between simulated density and experimental density was found in acetonitrile, but the difference remained under 5\% in all cases.

\subsection{Redlich-Kister Polynomial}

\subsubsection{Solvent Mixtures}
Once benchmarking was completed, simulations of various mixtures were performed. Fourteen solvent mixtures were simulated: a linear-linear carbonate mixture (DEC/DMC), a cyclic-cyclic carbonate mixture (EC/PC), cyclic-linear carbonate mixtures (EC/DMC, EC/DEC, PC/DEC, PC/DMC), carbonate-nitrile mixtures (EC/ACN, PC/ACN, DEC/ACN, DMC/ACN), carbonate-ether mixtures (EC/DME, PC/DME, DME/DMC), and an nitrile-ether mixture (DME/ACN). The excess densities were calculated from the simulation according to Equation \ref{eq:excess_eq} and fit to a Redlich-Kister polynomial according to Equation \ref{eq:RK_eq}. 

Fitting the excess densities of the solvent mixtures to a Redlich-Kister polynomial resulted in the coefficients shown in Table \ref{table:RK}. The magnitude and sign of the coefficient $a_1$ correlate to the magnitude and sign of the excess density. The second-degree Redlich-Kister polynomial is able to achieve a good fit, with an average SSE (sum of squared error) of 1.36E-08 $kg/m^3$. The mixture consisting of PC and ACN has the largest magnitude, and similarly has the largest value of $a_1$. Mixtures with a negative excess density have a positive $a_1$.  The coefficient $a_2$ correlates to the skew; the sign of $a_2$ is related to the direction of the skew of the curve. Although switching which solvent is $x_1$ and which is $x_2$ does not change the values of the coefficients, it can change the sign of $a_2$. Redlich-Kister fittings for experimental values for PC/DMC, PC/DEC, and DEC/DMC taken from literature \cite{francesconi1995excess} and measurements for PC/ACN taken with the robotic electrolyte testing platform Clio,\cite{dave2022autonomous} are also shown in Table \ref{table:RK}. 

\begin{table}[htb!]
\centering
\begin{tabular}{ |c|c|c|c| } 
 \hline
 $x_1/x_2$ & $a_1$ & $a_2$ & $a_3$ \\
 \hline
 EC/DME & 130.1  &	26.84  & 2.04  \\ 
 \hline
 EC/DMC & 28.18 & 4.72 & 0.56  \\ 
 \hline
 EC/DEC &  161.8  & 39.28 & -0.02934 \\ 
 \hline
 EC/PC & 14.3 & -0.6133 & -4.187  \\ 
 \hline
 PC/DME	& 30.18 &	11.09 &	-0.02133  \\ 
 \hline
 PC/DMC & -13.82 &	4.827  & -0.72  \\ 
 \hline
 PC/DEC	& 73.97 &	15.34  & 1.349  \\ 
 \hline
 DEC/DMC & 37.71  & -8.544  & 3.243  \\ 
 \hline
 DME/DMC & 42.61 & -4.741  & -0.1013  \\ 
 \hline
 PC/ACN	& -280.2 &	75.63 &	-11.09 \\ 
 \hline
 EC/ACN & -221.3  &	31.75  & 1.267  \\ 
 \hline
 DEC/ACN & -185 & 71.61  & -28.86 \\ 
 \hline
 DMC/ACN & -164.5  & -35.21  & -6.872  \\ 
 \hline
 DME/ACN & -80.46 & 26.27  & -3.752 \\  
 \hline
 Exp. PC/DMC & -19.92  & 6.971 & -1.917 \\
 \hline
 Exp. PC/DEC &  60.73  & 11.7  & 1.246 \\
 \hline
 Exp. DEC/DMC & 37.72 & -7.209 & 1.863 \\
 \hline
 Exp. PC/ACN & -244.8  & 73.46  & 10.96  \\
 \hline

\end{tabular}
\caption{Redlich-Kister polynomial coefficients for solvent mixtures.}
\label{table:RK}
\end{table}

\begin{figure}[htb!]
\centering
    \begin{subfigure}{0.49\columnwidth}
    \includegraphics[width=\columnwidth]{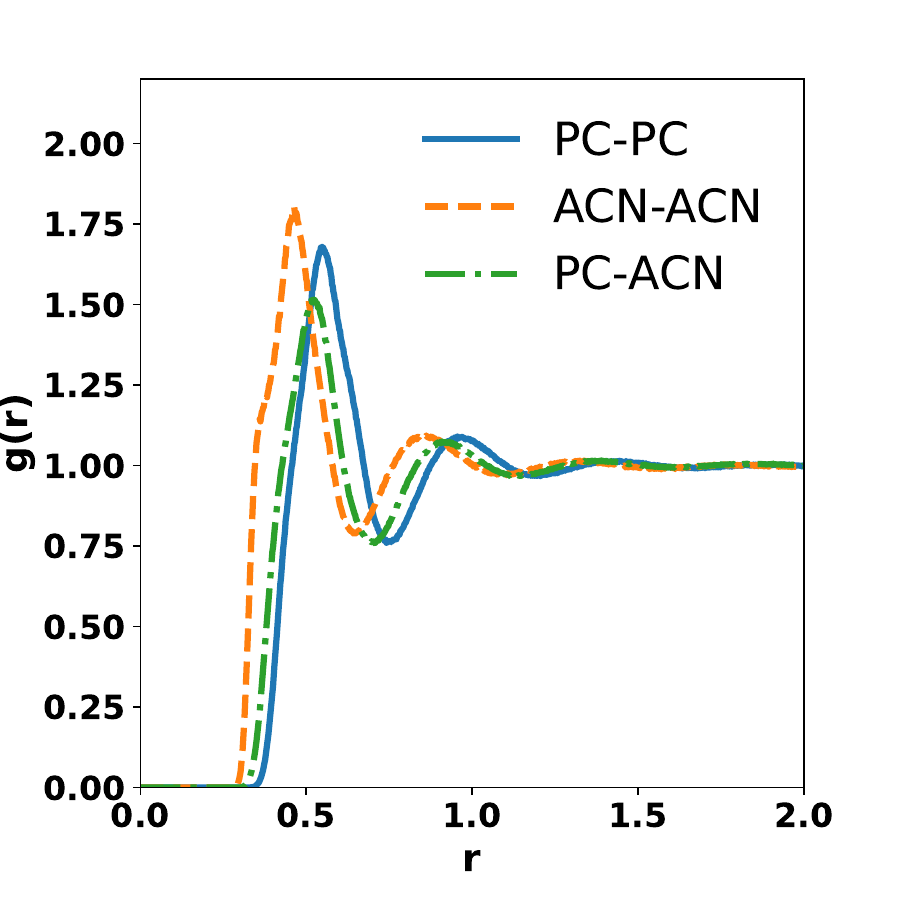}
    \subcaption{}
    \label{fig:rdfPCACN}
    \end{subfigure}
    %\hfill
    \begin{subfigure}{0.49\columnwidth}
    \includegraphics[width=\columnwidth]{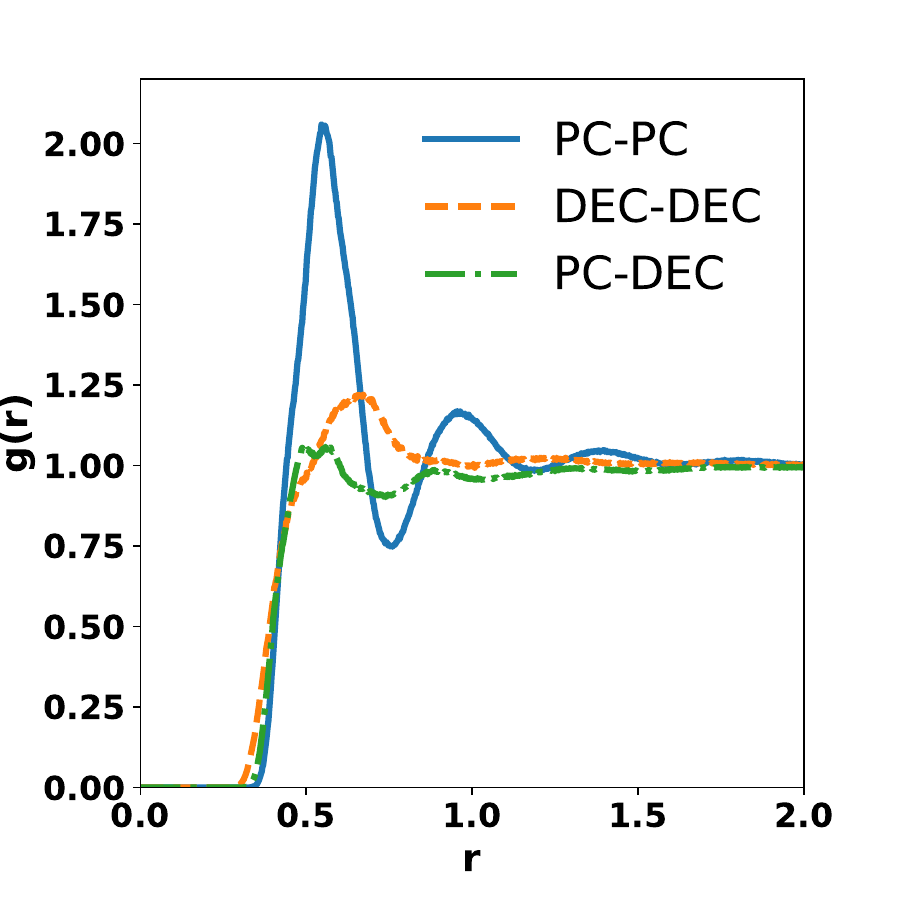}
    \subcaption{}
    \label{fig:rdfPCDEC}
    \end{subfigure}
\hfill
\caption{(a) shows the RDF of PC around PC, ACN around PC, and ACN around ACN in a mixture of 1:1 PC:ACN. (b) shows the RDF of PC around PC, DEC around PC, and DEC around DEC in a mixture of 1:1 PC:DEC}
\label{fig:rdf}
\end{figure}

The acetonitrile mixtures, PC/ACN, EC/ACN, DME/ACN, DMC/ACN, and DEC/ACN, experience positive excess densities, meaning that they are more dense than expected according to linear mixing. A potential cause could be geometric or structural factors. A comparison of the radial distribution function (RDF) of PC/DEC and PC/ACN in Figure \ref{fig:rdf} shows the RDF of ACN around ACN and the RDF of ACN around PC have strong peaks, indicating a more ordered structure. However, the RDF of DEC around PC and DEC around DEC have less intense peaks, indicating a less ordered structure. This more ordered structure may indicate that the ACN molecules are able to pack more closely than the ideal, leading to a highly positive excess density. 

\subsubsection{Solvent and Salt Mixtures}
\begin{figure}[htb!]
\centering
\includegraphics[width=\linewidth]{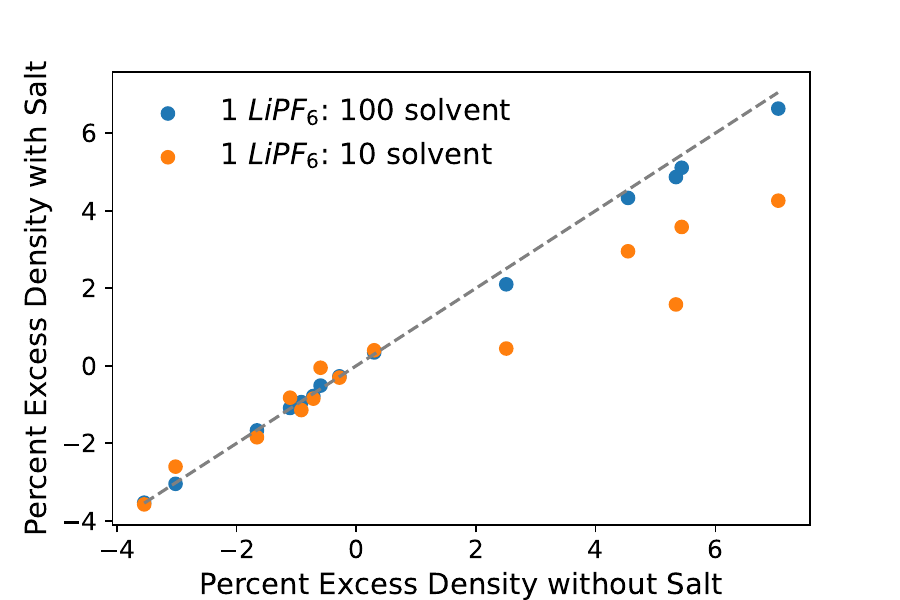}
\caption{The magnitude of the maximum percent excess density of solvent and salt mixtures is plotted against the magnitude of the maximum percent excess density of solvent mixtures. Two ratios of salt to solvent are shown, 1:100 in blue and 1:10 in orange.} 
\label{fig:comparison}
\end{figure}

Additional simulations were performed with these solvent mixtures and LiPF$_6$ in salt molecule to solvent molecule ratios of 1:100 and 1:10. The excess densities for the solvent and salt mixtures were fitted to Redlich-Kister polynomials, as shown in Table \ref{table:salt_RK} and Table \ref{table:salt_RK_10}. By comparing the first coefficients of the solvent mixtures and the solvent and salt mixtures, we can compare the magnitude of the excess density curves. This comparison is visualized in Figure \ref{fig:comparison}, which compares the maximum percent excess density with salt and without salt. 

The magnitude of the excess density does not uniformly increase or decrease across the mixtures when salt is added. For the 1:100 case compared to the solvent case, EC/DME, EC/DEC, PC/DME, PC/DMC, PC/DEC, and DEC/DMC all experience an increase in the magnitude of excess density, indicating an increase in the non-ideality of the mixture. EC/DMC, EC/PC, DME/DMC, PC/ACN, EC/ACN, DEC/ACN, DMC/ACN, and DME/ACN all experience decreases in the magnitude of excess density, indicating a decrease in the non-ideality of the mixture. The mixtures containing acetonitrile in particular experience a large decrease in excess density when salt ions are added. For the 1:10 case compared to the solvent case, EC/DEC, EC/PC, PC/DME, PC/DMC, PC/DEC, and DEC/DMC experience an increase in the magnitude of excess density, while EC/DME, EC/DMC, PC/ACN, EC/ACN, DEC/ACN, DMC/ACN, and DME/ACN all experience a decrease in the magnitude of excess density. The decrease observed in the acetonitrile mixtures increases further when the concentration of salt increases. These changes in percent excess density after salt is added compared to the percent excess density without salt are visualized in Figure \ref{fig:comparison}. 

Such changes in the behavior of excess density may be related to intermolecular forces or geometric factors. For the mixtures containing ACN, which have a positive excess density which decreases upon the addition of salt, this could reflect a relative decrease in attractive forces between the molecules compared to the solvent without salt. Mixtures with a positive excess density that increases in magnitude when salt is added could be experiencing an increase in attractive forces. Mixtures which have a negative excess density that decreases in magnitude could be experiencing a relative decrease in repulsive forces acting on the molecules, so that the molecules are able to pack more closely together and achieve a density closer to the ideal. Mixtures with a negative excess density that increases in magnitude, becoming further from linear mixing, could reflect a relative increase in repulsive forces such that the molecules are less closely packed together. 

The formation of solvation shells in the mixture may have an effect on the excess density, as the molecules will form solvation shells around the lithium ions. This change in structure between the solvent and the solvent and salt mixture could affect the difference in density compared to the ideal density from linear mixing.

\begin{table}[htb!]
\centering
\begin{tabular}{ |c|c|c|c| } 
 \hline
 $x_1/x_2$ & $a_1$ & $a_2$ & $a_3$ \\
 \hline
 EC/DME & 132.7	& 25.65	& -5.533  \\ 
 \hline
 EC/DMC & 24.4 & 4.747 & 1.387  \\ 
 \hline
 EC/DEC &  162.7 & 39.18 & 8.179 \\ 
 \hline
 EC/PC & 14	& -1.867 & -1.493  \\ 
 \hline
 PC/DME	& 33.17 & 7.933	& -1.315 \\ 
 \hline
 PC/DMC & -15.96 & 6.187 & -5.173 \\ 
 \hline
 PC/DEC	& 74.94 & 13.53 & 1.781  \\ 
 \hline
 DEC/DMC & 38.94 & -7.552 & 1.461  \\ 
 \hline
 DME/DMC & 42.52 & -6.947 & -0.5893 \\ 
 \hline
 PC/ACN	& -268.4 & 69.07 & -9.251 \\ 
 \hline
 EC/ACN & -211.8 & 29.27 & -3.483 \\ 
 \hline
 DEC/ACN & -172 & 64.46 & -24.49 \\ 
 \hline
 DMC/ACN & -160.2 & 33.96 & -5.096  \\ 
 \hline
 DME/ACN & -69.08 & 18.41 & -7.536 \\  
 \hline

\end{tabular}
\caption{Redlich-Kister polynomial coefficients for solvent and salt mixtures, with a ratio of 1 salt molecule to 100 solvent molecules.}
\label{table:salt_RK}
\end{table}

\begin{table}[htb!]
\centering
\begin{tabular}{ |c|c|c|c| } 
 \hline
 $x_1/x_2$ & $a_1$ & $a_2$ & $a_3$ \\
 \hline
 EC/DME & 123.7	& 28.8	& -12.53  \\ 
 \hline
 EC/DMC & 0.16 & 5.867 & 2.56  \\ 
 \hline
 EC/DEC &  176.9 & 22.53 & -0.3467 \\ 
 \hline
 EC/PC & 16.78	& 3.28 & 3.707  \\ 
 \hline
 PC/DME	& 39.24 & 8.325	& -13.18 \\ 
 \hline
 PC/DMC & -20.16 & 6.773 & -8.533 \\ 
 \hline
 PC/DEC	& 88.74 & -3.92 & 4.933  \\ 
 \hline
 DEC/DMC & 51.58 & 2.587 & 13.41  \\ 
 \hline
 DME/DMC & 35.64 & 3.355 & -9.872 \\ 
 \hline
 PC/ACN	& -196.7 & 45.04 & -4.456 \\ 
 \hline
 EC/ACN & -169.9 & 18.88 & -1.92 \\ 
 \hline
 DEC/ACN & -64.56 & 36.75 & 10.02 \\ 
 \hline
 DMC/ACN & -127.5 & 8.883 & -1.949  \\ 
 \hline
 DME/ACN & -17.31 & -1.731 & 4.243 \\  
 \hline

\end{tabular}
\caption{Redlich-Kister polynomial coefficients for solvent and salt mixtures, with a ratio of 1 salt molecule to 10 solvent molecules.}
\label{table:salt_RK_10}
\end{table}

\subsection{Descriptor for Excess Density}

\begin{figure}[htb!]
\centering
\begin{subfigure}{\columnwidth}
    \includegraphics[width=1\columnwidth]{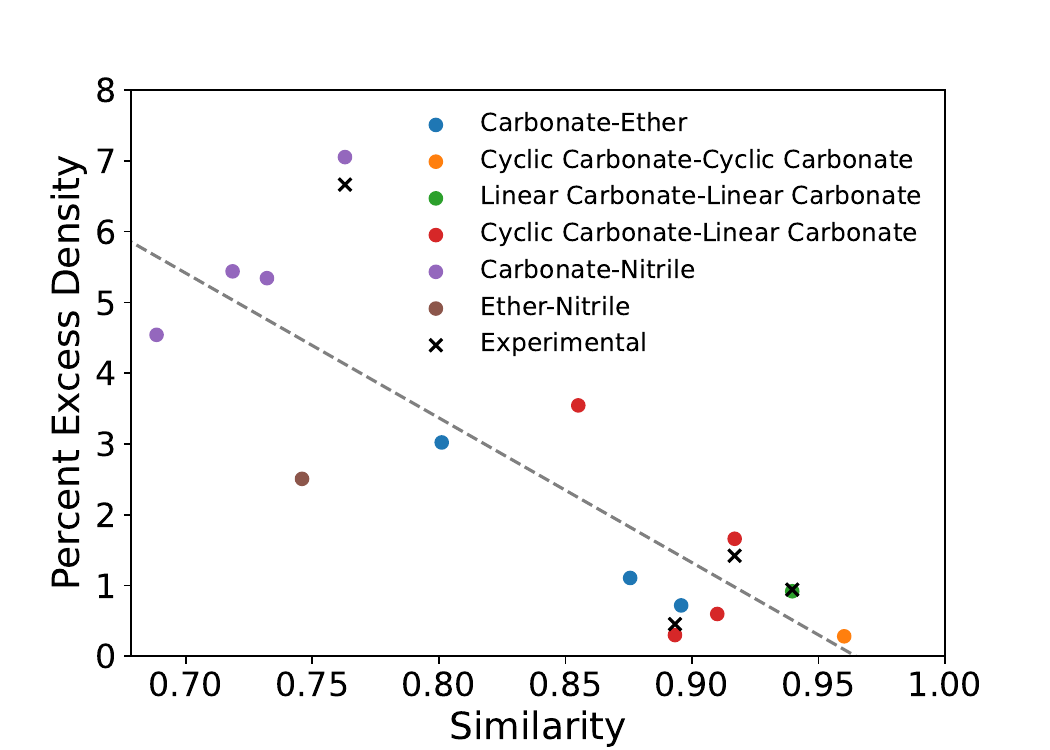}
    \caption{}
    \label{fig:similarity}
\end{subfigure}
\hfill
\begin{subfigure}{\columnwidth}
    \includegraphics[width=1\columnwidth]{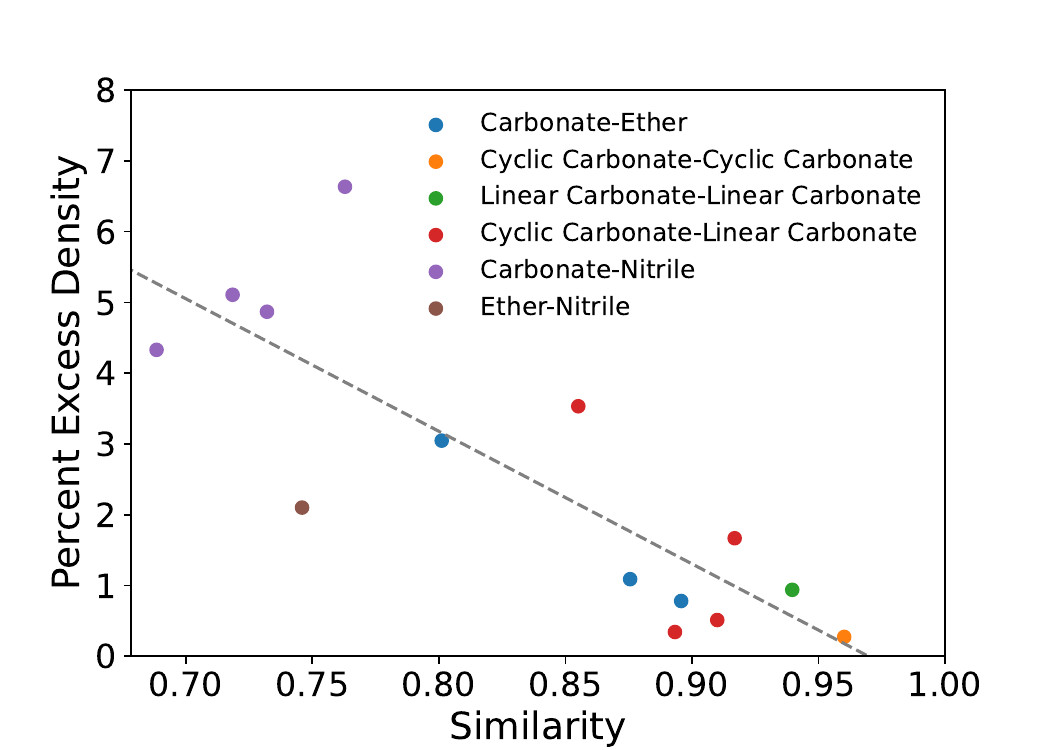}
    \caption{}
    \label{fig:similaritysalt}
\end{subfigure}
\hfill
\caption{(a) depicts solvent mixtures; (b) depicts solvent and salt mixtures. For both plots, the magnitude of the maximum percent excess density of each mixture is plotted against the similarity of the two components of the mixture. Carbonate-nitriles include EC/ACN, PC/ACN, DMC/ACN, and DEC/ACN. Ether-nitriles include DME/ACN. Carbonate-ethers include EC/DME, PC/DME, and DMC/DME. Cyclic-linear carbonates include EC/DMC, PC/DMC, EC/DEC, and PC/DEC. Linear-linear carbonates include DMC/DEC. Cyclic-cyclic carbonates include EC/PC. A linear fitting of the simulated results is shown with a grey dashed line. Experimental results are shown with black crosses on (a), vertically aligned with the corresponding simulation result.} 
\label{fig:similaritysaltnosalt}
\end{figure}

The similarities of the components of each mixture were found using SOAP fingerprints and the REMatch Kernel method. The resulting similarities can be seen in the leftmost plot in Figure \ref{fig:similarity}, which plots mixtures according to the similarity of the two components of the mixture. Mixtures containing similar solvents have a similarity closer to 1; mixtures containing dissimilar solvents have a similarity closer to 0. The excess densities compared here were normalized as the magnitude of the percent difference from linear mixing. The signed values do not exhibit as strong a linear trend as the absolute values. It may be that although mixing more dissimilar molecules can increase the deviation from linear mixing behavior, whether this deviation is positive or negative is affected by other factors, such as the repulsive or attractive forces between the molecules or the specific structures of the molecules used in the mixture. 

For each mixture, the maximum value of excess density was plotted. Experimental results from Comelli and Francesconi\cite{francesconi1995excess} are shown for the mixtures of PC/DMC, PC/DEC, and DEC/DMC, while those for PC/ACN were measured with Clio.\cite{dave2022autonomous} 

%\begin{figure}[htb!]
%\centering
%\includegraphics[width=\linewidth]%{figures/similarity_combined.png}

As can be seen in Figure \ref{fig:similarity}, the mixtures show a linear trend such that percent excess density tends to decrease as similarity increases; fitting the points to a line resulted in slope of -20.44 with an $R^2$ of 0.7. This can be used as a heuristic method to determine the behavior of binary mixtures of solvents: mixtures containing similar molecules are more likely to behave according to linear mixing, while mixtures containing dissimilar molecules are more likely to experience deviations from linear mixing (in either the negative or positive direction).  

The magnitude of the percent excess density of the salt mixture with a ratio of 1:100 salt to solvent molecules is also plotted against similarity, shown in Figure \ref{fig:similaritysalt}. Fitting the data to a line results in a slope of -18.73 with an $R^2$ of 0.676. The magnitude of the excess density changes for certain mixtures more than others upon the addition of the salt ions, resulting in the points fitting less closely to a line. The differing changes in excess density may be due to different molecules experiencing different degrees of attraction to or repulsion from the salt ions. 
 
%Excess molar volume is found with the following equation \cite{doi:10.1021/acs.jced.5b00366}, 
%\begin{equation}
%    V_E = [(x_1M_1 + x_2M_2)/\rho_mix] - [(x_1M_1/\rho_1)+(x_2M_2/\rho_2)]
%\end{equation} 

\section{Conclusion}
In this work, we examined the excess densities of several solvent mixtures containing ethylene carbonate, propylene carbonate, dimethyl carbonate, diethyl carbonate, dimethoxyethane, and acetonitrile. Good agreement with experiment was shown for PC/DMC, PC/DEC, DMC/DEC, and PC/ACN mixtures. Furthermore, salt and solvent mixtures with lithium hexafluorophosphate added were also considered. We fit this data to the Redlich-Kister polynomial, such that the magnitude and sign of the coefficients indicates the magnitude and direction of the excess density. 

The magnitude of excess density is affected by intermolecular forces or geometric factors. For instance, a  positive excess density may indicate that molecules are able to pack closely together, or that attractive forces dominate the interactions between molecules; a negative excess density may indicate that molecules are not able to pack closely together, or that repulsive forces dominate the interactions. We observe that mixtures containing acetonitrile tend to experience particularly high positive excess densities compared to other mixtures, particularly those that contain two linear carbonate molecules or two cyclic carbonate molecules. This indicates that attractive forces dominate over repulsive forces in the acetonitrile mixtures.  

To examine how the similarity of molecules affects the magnitude of excess density, we establish a design rule using a similarity metric based on the SOAP fingerprints of two solvent molecules and the REMatch kernel method. In doing so, we observed that excess densities tend to increase in magnitude as the similarity between the molecules decreases, and decreases as the similarity between the molecules increases. Based on this, we can use the similarity of molecules as a heuristic to predict the magnitude of the deviation from density calculated according to linear mixing rules. This gives us insights into the deviation from non-ideality a mixture would experience due to molecular interactions simply from the similarity of two molecules. As this non-ideal behavior witnessed in density would extend to other properties, predicting the excess density could potentially lead to predicting the non-ideality of other properties of interest. Further work should be done into investigating potential correlations between excess density and deviations from non-ideality in other properties. 

\section{Acknowledgements}
This work used Bridges-2 at Pittsburgh Supercomputing Center through allocation CTS180061 from the Advanced Cyberinfrastructure Coordination Ecosystem: Services \& Support (ACCESS) program, which is supported by National Science Foundation grants \#2138259, \#2138286, \#2138307, \#2137603, and \#2138296.\cite{bridges2} This work also used the Arjuna computer cluster, which was funded through Carnegie Mellon College of Engineering and the departments of Mechanical Engineering, Electrical and Computer Engineering and Chemical Engineering. We acknowledge funding from Advanced Research Projects Agency-Energy (ARPA-E), U.S. Department of Energy, from the EVs4ALL Program under award number DE-AR0001728. We also thank Yumin Zhang for her guidance in running molecular dynamics simulations. 

\section{Author Declarations}
The authors declare no competing financial interests.

\section{Author Contributions}
V.V. and C.K contributed to the design of the research. C.K., E.A., and V.V. contributed to the conceptualization of similarity as a descriptor for excess density. C.K. performed the molecular dynamics simulations, analyzed the simulations.  E.A. provided advice on the simulations and analysis thereof. A.D. performed the experimental measurements and experimental data analysis. C.K., E.A., and V.V. co-wrote the manuscript.

\section{Data Availability Statement}
The data that support the findings of this study are available from the corresponding author upon reasonable request.

\clearpage
\bibliography{bibliography}

\end{document}

\title{Supplemental Information for Excess Density as a Descriptor for Electrolyte Solvent Design}

%%%%%%%%%%%%%%%%%%%%%%%%%AUTHORS%%%%%%%%%%%%%%%%%%%%%%%%%%%%%%%%%%%
\author{Celia Kelly}
\affiliation{%
Department of Mechanical Engineering, Carnegie Mellon University, Pittsburgh, Pennsylvania 15213
}

\author{Emil Annevelink}
\affiliation{%
Department of Mechanical Engineering, Carnegie Mellon University, Pittsburgh, Pennsylvania 15213
}

\author{Adarsh Dave}
\affiliation{%
Department of Mechanical Engineering, Carnegie Mellon University, Pittsburgh, Pennsylvania 15213
}

\author{Venkatasubramanian Viswanathan}%
\email{venkvis@umich.edu}
\affiliation{%
Department of Mechanical Engineering, Carnegie Mellon University, Pittsburgh, Pennsylvania 15213
}
\affiliation{Department of Mechanical Engineering, University of Michigan, Ann Arbor, MI 48109}
\affiliation{Department of Aerospace Engineering, University of Michigan, Ann Arbor, MI 48109}
%\preprint{APS/123-QED}

\maketitle

\begin{figure}[htb!]
\centering
\includegraphics[width=\textwidth]{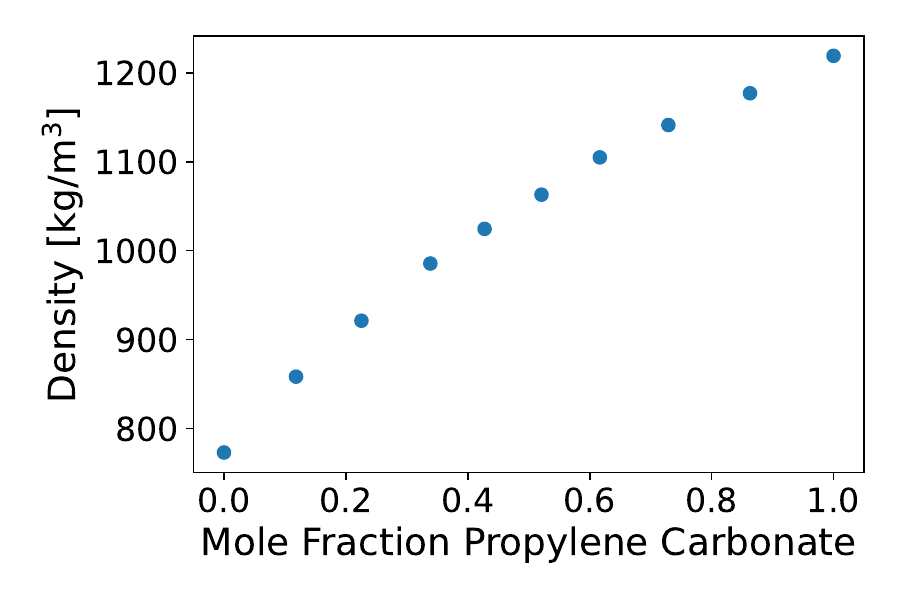}
\caption{Experimental results for a mixture of propylene carbonate and acetonitrile, measured with our robotic test stand, Clio.} 
\label{fig:experimental}
\end{figure} 

\begin{figure}[htb!]
\centering
\includegraphics[width=\textwidth]{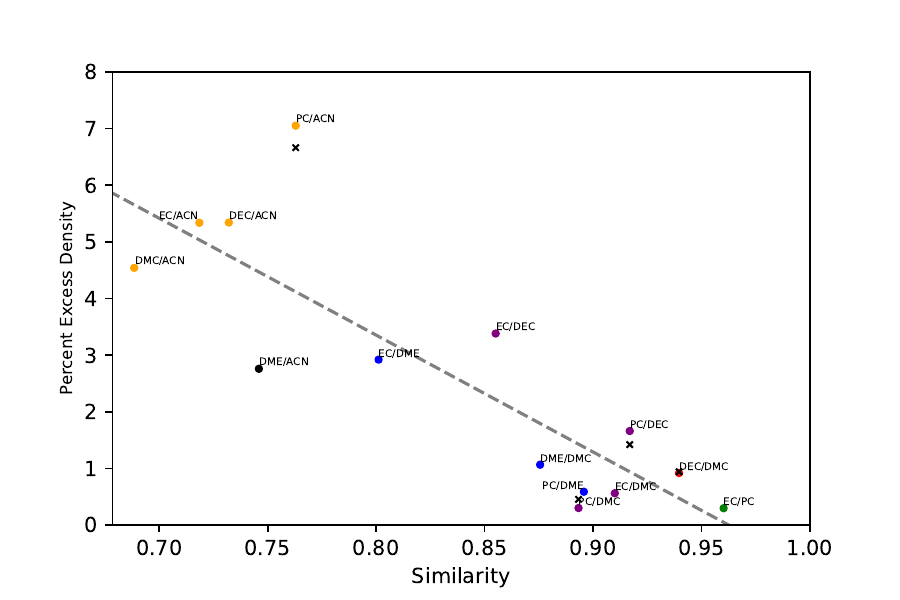}
\caption{The magnitude of the maximum percent excess density of each solvent mixture is plotted against the similarity of the two components of the mixture, with each point labeled. Carbonate-nitriles include EC/ACN, PC/ACN, DMC/ACN, and DEC/ACN. Ether-nitriles include DME/ACN. Carbonate-ethers include EC/DME, PC/DME, and DMC/DME. Cyclic-linear carbonates include EC/DMC, PC/DMC, EC/DEC, and PC/DEC. Linear-linear carbonates include DMC/DEC. Cyclic-cyclic carbonates include EC/PC. A linear fitting of the simulated results is shown with a grey dashed line. Experimental results are shown with black crosses.} 
\label{fig:SI_similarity}
\end{figure} 

\begin{figure}[htb!]
\centering
\includegraphics[width=\textwidth]{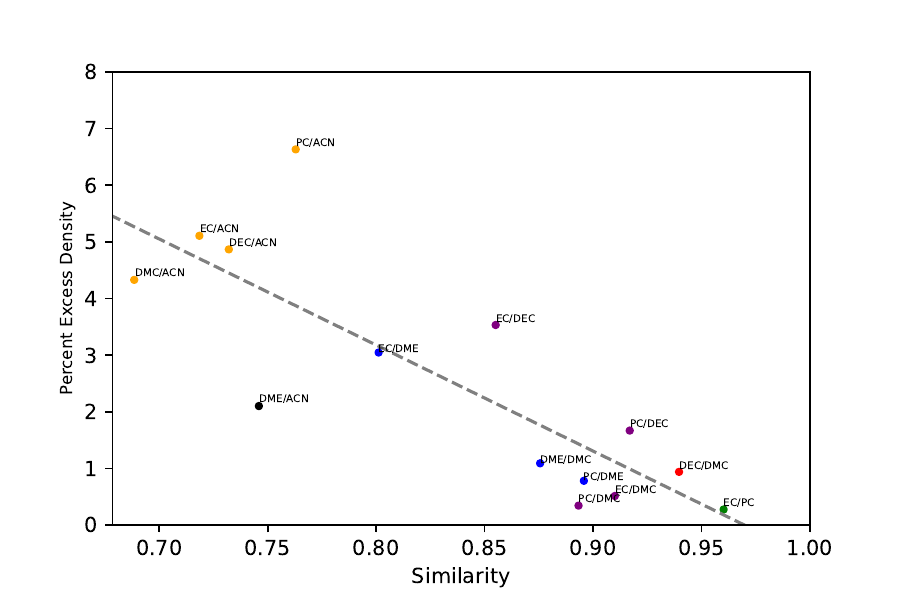}
\caption{The magnitude of the maximum percent excess density of each mixture of salt and solvents is plotted against the similarity of the two solvents in the mixture, with each point labeled. Carbonate-nitriles include EC/ACN, PC/ACN, DMC/ACN, and DEC/ACN. Ether-nitriles include DME/ACN. Carbonate-ethers include EC/DME, PC/DME, and DMC/DME. Cyclic-linear carbonates include EC/DMC, PC/DMC, EC/DEC, and PC/DEC. Linear-linear carbonates include DMC/DEC. Cyclic-cyclic carbonates include EC/PC.} 
\label{fig:SI_similarity_salt}
\end{figure}